\documentclass[pra,twocolumn,floats,showpacs,preprintnumbers,tighten,epsfig,superscriptaddress]{revtex4}

\usepackage{graphicx}
\usepackage{dcolumn}
\usepackage{bm}
\usepackage{amsmath}
\usepackage{amsfonts}
\usepackage{amssymb}

\usepackage{bm,color}

\begin{document}

\title{The Relativity Concept Inventory: development, analysis and results}

\author{J.~S.~Aslanides and C.~M.~Savage}
\email{craig.savage@anu.edu.au}
\affiliation{
Physics Education Centre, Research School of Physics and Engineering, Australian National University, Canberra ACT 0200, Australia}

\date{\today {}}

\begin{abstract}
We report on a concept inventory for special relativity: the development process, data analysis methods, and results from an introductory relativity class.  The Relativity Concept Inventory tests understanding of kinematic relativistic concepts. An unusual feature is confidence testing for each question. This can provide additional information; for example high confidence correlated with incorrect answers suggests a misconception. A novel aspect of our data analysis is the use of Monte Carlo simulations to determine the significance of correlations. This approach is particularly useful for small sample sizes, such as ours.  Our results include a gender bias that was not present in other assessment, similar to that reported for the Force Concept Inventory. 
\end{abstract}

\maketitle
\preprint{Version: Draft \today }

\section{Introduction}

Concept inventories are used to assess learning in many areas of physics education \cite{Inventory list}. When used to determine the effectiveness of educational innovations they may contribute to the teaching development cycle.   Since the literature on special relativity education research does not include a concept inventory we have developed the Relativity Concept Inventory (RCI), available from the Supplemental Appendix to this paper. 

Special relativity is interesting in a physics education research context because of its combination of deeply challenging concepts and  simple mathematics. This is in contrast with quantum mechanics, which has a more complex mathematical structure. Nevertheless, the amount of physics education research on special relativity is small \cite{Hewson 1982, Posner 1982, Villani 1987, Mermin 1994, Scherr 2001, Scherr 2002, Belloni 2004, Scherr 2007, Savage, McGrath, Wegener}. 

The RCI has been validated by feedback from discipline experts and its validity and reliability established by standard methods \cite{Chasteen, Adams}. These include the self-referential statistics of classical test theory, and benchmarking against traditional assessment such as homework and an exam. We have also developed and applied Monte Carlo simulation techniques suitable for the analysis of correlations in data with small sample size.

In the next section we describe the process used to develop the RCI. In section \ref{The Students} we characterize the students the RCI was administered to. In section \ref{Data Analysis Methods} we describe the methods used to analyse the collected data, including the use of: item response theory to control for the effect of student ability on correlations between questions, and Monte Carlo modeling to estimate the statistical significance of correlations. In section \ref{Results} we present  misconceptions diagnosed by the RCI and evidence for its gender bias. Finally, in the Conclusions, we suggest revisions of the RCI. We also argue that understanding the gender bias in concept inventories is a significant problem for physics education research.

\section{The Development Process}
\label{The Development Process}

The development of the RCI followed Adams and Wieman \cite{Adams} insofar as our six month project schedule allowed. In particular, student interviews were not relied on as much as suggested by them. The only previous attempt to develop a concept inventory for special relativity is reported in Gibson's doctoral thesis \cite{Gibson 2008}.

We first formulated a list of concepts that captured the learning goals of the introductory relativity instruction in the Physics 2 course at the The Australian National University (ANU).  These concepts were also informed by relevant textbooks \cite{Texts} and the physics education research literature.

Expert feedback on each of fourteen draft concepts of introductory relativity was obtained from thirty international respondents \cite{experts} using an online survey.  Agreement with the the appropriateness of the concepts in our list ranged from 100\% for the first postulate to 50\%. After individual consideration, concepts with agreement below 75\% were dropped from the list. The final list of nine concepts is given in Table \ref{concepts}. 

\begin{table*}
\caption{\label{concepts}The concepts tested by the RCI. In the questions column are the question number we classified as associated with each concept. Although some questions clearly test more than one concept we have allocated each question to only one concept.}
\begin{ruledtabular}
\begin{tabular}{p{4 cm}  p{11 cm} p{2 cm} }
  Concept & Description & Questions \\ \hline
First postulate. & The laws of physics are the same in all inertial reference frames. & 16, 18, 19 , 20 \\

Second postulate. & The speed of light in a vacuum is the same in all reference frames. & 3, 4 \\

Time dilation. &The time interval between two time-like separated events is shortest in the reference frame for which the two events are at the same position. The time between these events is greater in all other frames. & 5, 6, 7, 8 \\

Length contraction. & The length of an object (defined as the space interval between two simultaneous events at either end of the object) is the longest in the frame in which the ends of the object are at rest, and is shorter in all other frames. & 13, 14, 17 \\

\raggedright Relativity of simultaneity. & If two events A and B are space-like separated, then there exist inertial frames in which A precedes B, and others in which B precedes A. & 11, 12, 15, 21 \\

\raggedright Inertial reference frame. & A coordinate system in which a free particle will maintain constant velocity; in particular, the concept that all inertial frames are equivalent. & 1, 2 \\

Velocity addition. & Velocities transform between frames such that no object can be observed travelling faster than the speed of light in a vacuum.  & 9, 10 \\

Causality. & If two events are time-like separated, then the ordering of the events is fixed for all reference frames. &  22, 23 \\

\raggedright Mass energy equivalence. & Energy has inertia. & 24
\end{tabular}
\end{ruledtabular}
\end{table*}

These concepts were used to develop twenty-four draft RCI multiple-choice questions, with one, two or three questions primarily addressing each of the concepts.  Expert feedback on the draft RCI questions was obtained from seven respondents using another online survey. In addition, a face-to-face interview was conducted with the ANU academic teaching advanced special relativity.  

It was then administered to six fourth-year physics students. These students were also asked to write a sentence or two explaining their reasoning for each question. Next, the RCI was taken by three second-year students in think aloud format: students were asked to verbalise their thinking while answering the RCI questions.  These students had taken Physics 2 the previous year. These sessions were recorded and transcribed for study. 

The RCI was then administered online to the 2012 ANU Physics 2 class, prior to instruction, as a pre-test, and after instruction as a post-test. Neither contributed to the course assessment. Students' RCI post-test responses were compared to their answers to the relativity questions in the Physics 2 mid-course exam, which included short answer conceptual questions.

All this feedback was used to continuously improve the draft RCI. Wording was clarified when found to be ambiguous and questions were deleted when it was determined they were not adequately addressing desired concepts. The final version of the RCI is available from the Supplemental Appendix to this paper.  It consists of twenty-four multiple choice questions, with each having a confidence scale.  Example questions are given in Table \ref{questions}. Throughout this paper individual questions are referred to by their RCI question number.

\begin{table}
\caption{\label{questions} Questions 5, 6, 7 and 23 from the RCI. The first three test the time dilation concept. The correct answer to each is (a). Question 23 tests multiple concepts. The correct answer is (d). The full RCI may be found in the supplemental appendix.}
\begin{ruledtabular}
\begin{tabular}{p{8.5 cm} }

\textit{In the following two questions, Abbey is in a spaceship moving at high speed relative to Brendan, who is standing on an asteroid (a very small rock floating in space). She flies past him so that at $t=0$, she is momentarily adjacent to Brendan.}
\vspace{0.1 cm}

5. At the instant that Abbey's ship passes Brendan, she sends two light pulses to him from her ship. If the light pulses are emitted a nanosecond ($10^{-9}$ seconds) apart according to Abbey's clock, what will be the time interval between the pulses according to Brendan?
\vspace{0.1 cm} \\

(a) Greater than one nanosecond \\

(b) Equal to one nanosecond \\

(c) Less than one nanosecond \\

 \hline
\\
6. Also while Abbey's ship passes Brendan, Brendan sends two light pulses to Abbey.  If Brendan sends the light pulses a nanosecond ($10^{-9}$ seconds) apart according to his clock, what will be the time interval between the pulses according to Abbey?
\vspace{0.1 cm} \\

(a) Greater than one nanosecond \\

(b) Equal to one nanosecond \\

(c) Less than one nanosecond\\
\hline
 \\
7. It is known that our galaxy is of the order of 100,000 light-years in diameter. True or false: ``Travelling at a constant speed that is less than, but close to, the speed of light, in principle it is possible for a person to cross the galaxy within their lifetime.''
\vspace{0.1 cm} \\

(a) True \\

(b) False. \\

 \hline
 \\
23. If two events are separated in such a way that \textbf{no} observer can be present at both events, which relationship(s) are the same for all observers?
\vspace{0.1 cm} \\

(a) The time between the two events \\

(b) The distance between the two events \\

(c) The order in which the events occur \\

(d) None of these relationships are the same for all observers
\end{tabular}
\end{ruledtabular}
\end{table}

RCI questions have an associated confidence scale which asks the student to rate how confident they are in their answer. One of five options could be selected from the online form: guessing, unconfident, neutral, confident, and certain. Confidence measures have occasionally been used before with concept inventories \cite{Allen, Allen 2004, Allen 2006}, including in association with the FCI \cite{Sharma}.   

Confidence information is potentially useful for gauging the quality of students' understanding. For example, consider a question that most students answer correctly. If they also expressed confidence in their answers this would suggest mastery had been achieved. This was the case for the pair of questions 3 and 4 concerning the constancy of the speed of light. However if students expressed less confidence it might indicate memorisation or shallow understanding. This was the case for the pair of questions 5 and 6 concerning time dilation, see Table \ref{questions}. 

Perhaps more interesting are questions that are answered incorrectly for which students indicate confidence in their answer. This indicates a potential misconception. This was the case for question 7 concerning a twin paradox type scenario; see Table \ref{questions}.

\section{The Students} 
\label{The Students}

The RCI data analyzed in the rest of this paper was obtained from the 2012 ANU Physics 2 class \cite{ethics}. This is the second physics course taken by physics majors.  The class enrolment was niety-nine, from whom seventy responses were obtained for the pre-test and sixty-three responses for the post-test, with fifty-three individuals taking both tests. 

The relativity instruction was a three week module of: nine lectures, a three hour simulation laboratory using the Real Time Relativity software \cite{RTR}, and three small-group problem-solving tutorials. It was assessed by two sets of weekly homework, a pre-lab problem, a lab log-book, and a mid-term exam question. The lectures were held in a studio space to encourage interaction, and included clicker questions and small group discussion.

The RCI was administered online in 30 minutes of scheduled class time, although those absent from class were able to complete it outside of class time. No significant differences were found between those two groups. All questions were of equal value, with no partial marks given. The mean RCI score on the pre-test was 56\%, and on the post-test 71\%. For comparison, the expected mean score if answers were chosen randomly is 36\%, with a standard deviation of about 1\% (see section \ref{Monte Carlo simulation} for further explanation). These high scores should be considered in the context of the class being high academic achievers, as indicated by their median Australian Tertiary Admission Rank (ATAR) score of 95, out of a possible 99.95 \cite{ATAR}. 

For our analysis we numerically coded the five confidence options as: guessing (0), unconfident (0.25), neutral (0.5), confident (0.75) and certain (1). The mean confidence over all questions and all students was then 0.5 for the pre-test and 0.68 for the post-test. The average of the Pearson correlation, Eq.~(\ref{Pearson correlation}), between students' confidence and their score for each question was $\langle r_i \rangle = 0.11$ for the pre-test and $\langle r_i \rangle = 0.19$ for the post-test. Hence, after instruction students not only became more confident but were also more likely to answer correctly if they expressed confidence.

Interestingly, although approximately a third of the class claimed to have had prior formal instruction in relativity at secondary school, those students did not perform better in either the RCI pre or post-tests, or in the exam relativity question.

\section{Data Analysis Methods} 
\label{Data Analysis Methods}

In this section we analyse the data obtained from administering the RCI to the Physics 2 class. In section \ref{Classical test theory} we use classical test theory to investigate the discrimination and consistency of the RCI. In section \ref{Question correlations} we investigate the correlations between students' responses to different RCI questions.

As our sample size is small we paid particular attention to the statistical significance of correlations. Where possible, we calculated the probability that the observed correlations might arise by chance from sampling noise rather than from actual properties of the underlying population: so called $p$-values. In the language of physics and engineering, we attempted to distinguish the signal from the noise \cite{Silver}. 

In the case of approximately normally distributed data this was done using standard deviations from the mean. Otherwise, we used either the Kolmogorov-Smirnov test \cite {numerical recipes}, or Monte Carlo simulations, to calculate the probability that the correlation could have arisen by chance. The Kolmogorov-Smirnov test is preferred over the chi-squared test for small sample sizes \cite{Wackerly}.

\subsection{Classical test theory}
\label{Classical test theory}

Classical test theory provides a set of statistics for estimating the discrimination and consistency of a test. Discrimination is the capability to quantify students' understanding of the subject of the inventory. Consistency is the extent to which each question is measuring the same broad understanding. Overviews have been given by Ding \textit{et al.} \cite{Ding2006}, and Ding and Beichner \cite{Ding}. 

\begin{figure}[tbp]
\includegraphics[width=8cm]{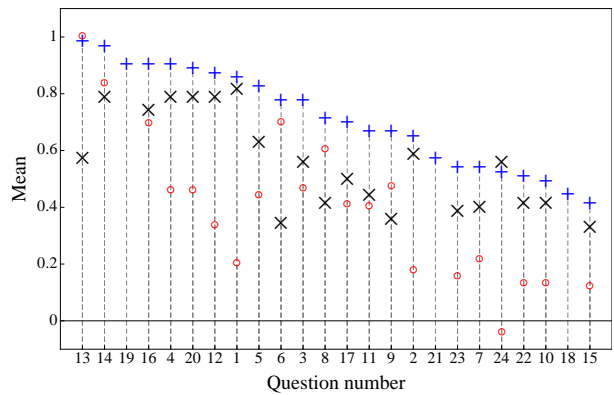}
\caption{(colour online) RCI results by question for the Physics 2 class: the  post-instruction item difficulties (blue +),  pre-instruction item difficulties (black ${\times}$), and the normalised gain (red ${\circ}$).  The sample sizes were 63 for the post-test and 70 for the pre-test, with 53 individuals doing both tests. The question number ordering is by post-instruction item difficulty. Questions 18, 19 and 21 have no pre-test item difficulties or normalised gains as they were changed between the pre and post-tests. The actual post-test questions are given in the Supplemental Appendix. The normalised gain is calculated for the students who took both the pre-test and the post-test. Hence it cannot be calculated using the plotted pre and post scores, as they include additional students.}
\label{scores}
\end{figure}

Table \ref{RCI test stats} reports some test statistics for the RCI post-test. The desired ranges are boundaries, according to Ding and Beichner \cite{Ding}, beyond which consideration should be given to possible problems with the inventory. The item difficulty of question number $i$ is the fraction of correct answers, $P_i =N_{\textrm{correct}}/N_i$, where $N_i$ is the total number of answers to the question. Figure \ref{scores} shows the item difficulties for each question. The post-test RCI item difficulty averaged over all questions,  $\langle P \rangle = 0.71$, tells us that the test was rather easy. However, as noted in the previous section, the class was particularly accomplished.
\begin{table}
\caption{\label{RCI test stats}RCI post-test statistics. Sample size $N=63$ students. The desired ranges are those suggested by Ding and Beichner \cite{Ding}.}
\begin{ruledtabular}
\begin{tabular}{lcc}
  Statistic & RCI value & Desired range \\ \hline
Mean item difficulty & 0.71 & [0.3,0.9] \\
Mean discrimination index & 0.24 & $\ge$ 0.3 \\
Ferguson's delta & 0.96 & $\ge$ 0.9 \\
Mean point biserial coefficient & 0.36 & $\ge$ 0.2 \\
KR20 reliability & 0.74 & $\ge$ 0.7
\end{tabular}
\end{ruledtabular}
\end{table}
For those questions that did not change between the pre-test and post-test, Fig.~\ref{scores}  shows the pre-test item difficulties and the normalised gain. The normalised gain for a question is defined to be the change in item difficulty divided by the maximum possible change in item difficulty, $g_i =(P_{i,\textrm{post}} -P_{i,\textrm{pre}})/(1-P_{i,\textrm{pre}})$ \cite{Hake 98}.  It is the fraction of the possible improvement that was achieved following instruction. The RCI normalised gain averaged over all questions was $\langle g \rangle = 0.40$. The Kolmogorov-Smirnov test \cite {numerical recipes} determined that the probability that the pre and post-test results were sampled from the same population was $p=4 \times 10^{-6}$. Hence we conclude that the normalised gain is statistically significant.

The only RCI statistic in Table \ref{RCI test stats} falling outside the desired range is the mean discrimination index. This compares the number of students whose total RCI results were in the top quartile to those in the bottom quartile. The discrimination index for a question takes the difference between the fraction of correct answers to that question from students in the top quartile $N_{i,\textrm{T}}$ and from those in the bottom quartile $N_{i,\textrm{B}}$: $D_i = N_{i,\textrm{T}} /(0.25 N_i ) -N_{i,\textrm{B}} /(0.25 N_i )$. The mean discrimination index is the mean of the discrimination indices for all questions. The low RCI value in Table \ref{RCI test stats} is partially due to the ease of the RCI, discussed in section \ref{The Students}, which reduces discrimination because the difference in student performance between the top and bottom quartiles is less than for a difficult test. Questions 12, 13, 14, 20 and 24 had discrimination indices $D_i \le 0$. Their range of item difficulties was $0.52 \ge P_i \ge 0.98$ with a mean of 0.85. These questions should be reconsidered in any RCI revisions. Indeed, in section \ref{Item response theory} we recommend dropping question 24, concerning mass-energy equivalence. Hence, the low mean discrimination index suggests how the RCI might be improved. Nevertheless, we next show that another measure of discrimination, Ferguson's delta, is within the acceptable range.

Ferguson's delta measures how the actual total scores are distributed in comparison to the possible range of scores. If only one particular score was ever achieved then $\delta = 0$, while if all possible scores are achieved equally often $\delta \approx 1$. Thus Ferguson's delta measures the ability of the RCI to discriminate between students' understanding. It is defined to be \cite{Ding}
\begin{equation} \label{fergdelta}
\delta = \frac{N^2 -\sum_{i=1}^K f_i^2 }{N^2 -N^2/(K+1)}  ,
\end{equation}
where $f_i$ is the number of times the total score was $i$.  In contrast to the discrimination index, the RCI Ferguson's delta of $\delta = 0.96$ indicates that the RCI has adequate discrimination. We conclude that while the discrimination of the RCI might be improved, it is adequate.

The Pearson correlation between random variables $X$ and $Y$ is defined to be their covariance divided by the product of their standard deviations:
\begin{equation} \label{Pearson correlation}
r_{\textrm{XY}} = \frac{\textrm{Cov}(X,Y)}{\sqrt{\textrm{Var}(X) \textrm{Var}(Y)}} .
\end{equation}
where $\textrm{Cov}(X,Y) = \langle (X- \langle X \rangle ) (Y- \langle Y \rangle ) \rangle$ and $\textrm{Var}(X) = \langle (X- \langle X \rangle )^2 \rangle$. In classical test theory the point biserial coefficient for a question is the Pearson correlation between its item score and the total score for the inventory. Treating question answers as dichotomous variables, being right or wrong, the point biserial coefficient for question number $i$ can be expressed as  \cite{Ding}
\begin{equation} \label{pbsc}
r_{\textrm{pbc},i} = ( \langle X_{r,i} \rangle - \langle X_{w,i} \rangle ) \sqrt{P_i (1-P_i )} / \sigma_X  ,
\end{equation}
where $\langle X_{r,i} \rangle$ is the mean total score for those who got the question right, $\langle X_{w,i} \rangle$ is the mean total score for those who got the question wrong, and $\sigma_X$ is the standard deviation of the total score. The RCI mean point biserial coefficient over all post-test questions of $\langle r_{\textrm{pbc}} \rangle = 0.36 $ tells us that the RCI questions  are consistent in what they measure.

The KR20 reliability statistic is another measure of the internal consistency of the inventory. It estimates the degree of correlation between the answers to questions. A value near one indicates that all questions are testing the same thing, while a value near zero indicates that the answers are independent of each other. A value too close to one would be undesirable for the RCI, since it is intended to test a number of different concepts. However, as usual in physics, the concepts are interrelated, so that a deep understanding of relativity requires an understanding of all concepts; so a low value is also undesirable. The KR20 reliability statistic is defined to be \cite{Ding}
\begin{equation} \label{pbsc}
r_{\textrm{KR20}} = \frac{K}{K-1} ( \sigma_X^2 -\sum_{i=1}^K \sqrt{P_i (1-P_i )} \; ) / \sigma_X^2  ,
\end{equation}
where $K = 24$ is the number of questions in the inventory. The RCI reliability statistic of $r_{\textrm{KR20}} = 0.74$ agrees with the mean point biserial coefficient that the RCI questions are consistent in what they measure.

\subsection{Question correlations}
\label{Question correlations}

Correlations between students' responses to different questions can provide information on the reliability of the Inventory. They can also provide information about students' understanding, as we will show in section \ref{Misconceptions}. 

As usual in statistical analysis, we assume that our sample, the Physics 2 class, is a subset of a larger population that we want to understand. This might be all students who have taken, or will take, a similar course. We assume that our sample of students is randomly chosen from the larger population and that its statistics estimate those of the larger population. However, in the particular sample, correlations can arise by chance even when no underlying correlation exists. Hence it is important to calculate the statistical significance of correlations, especially with small sample sizes, such as ours. This tells us the probability that we might be misled by sample noise, and hence informs any action that might be taken based on the statistical evidence.

\begin{figure}[tbp]
\includegraphics[width=8cm]{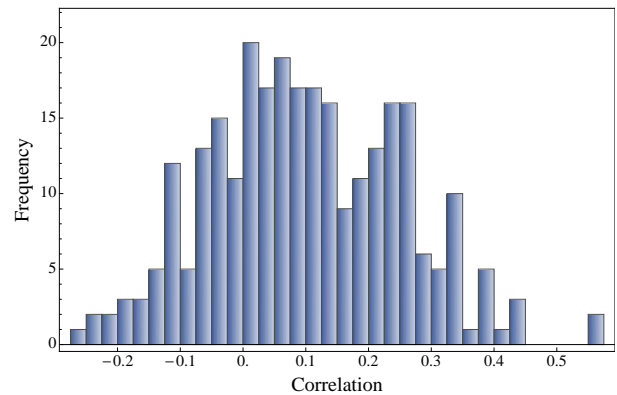}
\caption{(colour online) Histogram of the Pearson correlations between all 276 question pairs from the post-test data. The correlations are calculated using Eq.~(\ref{multinomial correlation}), with the $p_{XY}$ derived from the data. The mean correlation is $\langle r \rangle = 0.1$ and the standard deviation is $0.15$. }
\label{question correlations}
\end{figure}

For example, with twenty-four questions in the RCI there are $(24 \times 23) /2 = 276$ possible correlations between question pairs. These are shown in Fig.~\ref{question correlations}, as calculated from the post-test data. To understand why this should alter our choice of statistical significance threshold, assume there was a hypothetical 5\% chance of correlations above a certain strength occurring between any particular question pair, entirely due to random variation in the data. Then we would expect to find about $276 \times 0.05 \approx 14$ so correlated question pairs by chance. Choosing an acceptance threshold of $p < 1/276 \approx 4\times10^{-3}$ ensures that in the long run less than one correlation is accepted due to sampling noise alone. Such care is required whenever there are many noisy channels in which a signal is being sought. However, it comes at the cost of an increased likelihood of missing correlations that in fact exist in the larger population.

A related problem is determining the significance of the absence of expected correlations. For example, consider two questions that were designed to test the same concept, but that are not significantly correlated according to the student data. What strength of correlation can the data reliably rule out?

We have addressed such questions using Monte Carlo simulation. As this approach is not common in physics education research, we describe it in some detail in the next section.

\subsubsection{Monte Carlo simulation}
\label{Monte Carlo simulation}

Our Monte Carlo simulations are based on stochastic models of the student population. Random samples are drawn from the model  and their distributions used to estimate statistical significance. As models are simplified descriptions of students' responses, such estimates must be treated with care. Nevertheless, they help quantify the degree to which correlations in the data imply correlations in the larger population.

An example, concerning means rather than correlations, was given in section \ref{The Students}. The standard deviation in randomly answered mean scores was estimated from a model in which the answer to each question was chosen with uniform probability. The mean scores of samples of size $N=70$  were approximately normally distributed with a mean of 36\% and a standard deviation of about 1\%. Since the pre-test mean of 56\% is then about twenty standard deviations from the mean, we can conclude that the students are not guessing their answers.

More interesting is the estimation of the statistical significance of correlations between two questions. Let us call them Q1 and Q2. We code the question answers as correct (1) or incorrect (0). There are then four possible answers to the two questions: both correct, both incorrect, only Q1 correct, and only Q2  correct. Our model of the larger student population assumes that students' answers follow the multinomial distribution over these four possible outcomes. 

Let $p_{11}$ be the probability that both questions are answered correctly, $p_{00}$  the probability that both are answered incorrectly, $p_{10}$ the probability that only Q1 is answered correctly, and $p_{01}$  the probability that only Q2 is answered correctly. The multinomial probability function is then \cite{Wackerly}
\begin{align} \label{multinomial}
\textrm{Pr}(N_{11}, N_{00}, N_{10}, N_{01}) =& \frac{N!}{N_{11}! N_{00}! N_{10}! N_{01}!} \nonumber  \\
& \times p_{11}^{N_{11}} p_{00}^{N_{00}} p_{10}^{N_{10}} p_{01}^{N_{01}},
\end{align}
where $N_{XY}$ is the number of $XY$ outcomes from a sample of $N$ answers. Three equations, in addition to the normalization, $p_{11}+p_{00}+p_{10}+p_{01} = 1$, specify the distribution. We take these to be the probability of a correct answer to Q1, $P_1 = p_{11} +p_{10}$, the probability of a correct answer to Q2, $P_2 = p_{11} +p_{01}$, and the Pearson correlation between the answers to Q1 and Q2, 
\begin{align} \label{multinomial correlation}
r_{12} = \frac{p_{11} p_{00} - p_{10} p_{01}}
{ \sqrt{ (p_{11}+p_{10}) (p_{11}+p_{01}) (p_{00}+p_{10}) (p_{00}+p_{01}) } } .
\end{align}
Hence specifying $P_1$, $P_2$, and $r_{12}$ determines the distribution. The first two are the item difficulties from the student data. In contrast, the correlation is chosen to test a significance hypothesis. For example, say the student data has a correlation of $C$, and we want to know whether this is significant. We then choose the model correlation to be $r_{12} = 0$. Taking Monte Carlo samples from the model \cite{Mathematica} we can determine the probability that correlations equal to or larger than the observed correlation $C$ arise from the model with zero correlation. If this probability is $p$ we would say that the observed correlation is statistically significant at the $p$ level. 
\begin{table}[t]
\caption{\label{post-test correlations} Post-test correlations between questions statistically significant at the $p \le 10^{-3}$ level. The Pearson correlation is calculated using Eq.~\ref{multinomial correlation}. The $p$-values were obtained from $20,000$ Monte Carlo samples for each question pair with zero correlations between questions.}
\begin{ruledtabular}
\begin{tabular}{ccc}
Questions & Pearson's $r$ & $p$-value \\ \hline
1, 2 & 0.56 & $< 5\times10^{-5}$\\ 
5, 6 & 0.56 & $< 5\times10^{-5}$\\
11, 12 & 0.44 & $4\times10^{-4}$\\
3, 9 & 0.43 & $3\times10^{-4}$\\
15, 22 & 0.44 & $5\times10^{-4}$\\ 
2, 7 & 0.39 & $7\times10^{-4}$\\
9, 22 & 0.38 & $9\times10^{-4}$
\end{tabular}
\end{ruledtabular}
\end{table}

Monte Carlo significance testing of our post-test data found the seven correlations shown in Table \ref{post-test correlations} to be significant at the $p \le 10^{-3}$ level. From the argument at the beginning of section \ref{Question correlations} these are unlikely to arise randomly. The first three are expected correlations between conceptually related questions. However the others are unexpected. In the next section we explain the observed correlations between these conceptually unrelated questions using item response theory. 

It is surprising that Table \ref{post-test correlations} does not contain more correlations between conceptually related questions. However, the fact that an observed correlation is not statistically significant does not, in itself, justify the conclusion that there is no correlation in the larger population. As far as the data alone is concerned, it leaves us uncertain either way. 

One way of dealing with this problem is based on Bayes' theorem \cite{Silver}. In our context, this approach assigns prior probabilities to correlations. These probabilities are then adjusted according to the statistical evidence from the data. This has the advantage that correlations that we have prior reason to believe exist, for example between conceptually related RCI questions, are less likely to be rejected as noise than do correlations that we have no prior reason to believe exist. Although we will not use quantitative Bayesian statistics, the Bayesian framework helps explain the lack of expected correlations in Table \ref{post-test correlations}, as it takes no account of prior information.

Alternatively, further Monte Carlo simulations might show that sufficiently strong correlation values are unlikely.  In cases for which we expected a correlation, this would justify a reconsideration of our reasons for that expectation. For example, we could select an assumed strong correlation $C_A$ and set the model correlation equal to it, $r_{12} = C_A$. From Monte Carlo simulations we could then determine the probability $p$ that the simulated correlations are equal to or less than the observed correlation $C$, even though the model correlation is $C_A$. If this probability is sufficiently small we may rule out the assumed correlation at the $p$ level.

\subsubsection{Item response theory}
\label{Item response theory}

It is reasonable to assume that a major determinant of whether a student answers a question correctly is their academic ability. Given a question pair, strong students will tend to get both right, and weak students will tend to get both
wrong, strengthening the overall correlations. If this assumption is correct, then removing that part of students' performance due to academic ability may increase the correlations due to conceptual relations \cite{Francis}. This may be achieved using item response theory \cite{Ding}.

Item response theory, sometimes called Rasch analysis \cite{Planinic}, assumes that there is one parameter that describes
the performance of student number $j$, their ability $\theta_j$, and one parameter, $b_i$, that describes the difficulty of question number $i$. These are generated by a logistic regression algorithm \cite{ministeps} from the student data to provide a maximum likelihood estimate for the probability of student $j$ getting question $i$ correct from the model 
\begin{equation} \label{RT model}
P_{ij}= \frac{e^{\left(\theta_{j}-b_{i}\right)}}{1+e^{\left(\theta_{j}-b_{i}\right)}} .
\end{equation}
Let $M_{ij}$ be the actual response of student $j$ to question $i$, coded so $1$ is correct and $0$ incorrect. The residuals $R_{ij} = M_{ij} -P_{ij}$ measure the deviation of the particular student $j$ and question $i$ from the population of students and questions with the same respective ability and difficulty. According to item response theory these residuals have the student ability and question difficulty factors removed. Hence correlations between the residuals are due to factors other than student's ability and question difficulty.

We therefore calculated the correlations between the residuals for each question pair, averaged over all $N$ students
\begin{equation} \label{residual correlations}
C_{ik}={\displaystyle \frac{1}{N}\sum_{j=1}^{N}}R_{ij}R_{kj} .
\end{equation}
These correlations were found to be approximately normally distributed with mean zero and standard deviation $0.02$. We consider the statistically significant correlations to be those that are more than three standard deviations from the mean, that is, with a one-sided $p$-value of $< 2 \times 10^{-3}$. Table \ref{RT correlations} lists these. 
\begin{table}[t]
\caption{\label{RT correlations} Item response theory residual correlations $C_{ik}$, statistically significant at the three-sigma level, from the post-test data. The rightmost column is how many standard deviations $C_{ik}$ is from the mean.}
\begin{ruledtabular}
\begin{tabular}{ccc}
Questions & $C_{ik}$ & $\sigma$ \\ \hline
5, 6 & 0.08 & $4.0$\\ 
1, 2 & 0.066 & $3.4$ \\ 
11, 12 & 0.066 & $3.4$\\ 
7, 8 & -0.083 & $3.6$\\ 
23, 24 & -0.086 & $3.7$
\end{tabular}
\end{ruledtabular}
\end{table}

The three positively correlated questions are precisely the conceptually related pairs in the raw scores correlation Table \ref{post-test correlations}.  All the other correlations in Table \ref{post-test correlations} are absent. Hence student ability, as modelled by item response theory, explains the correlations between the raw scores of conceptually unrelated questions.

The last two rows in Table \ref{RT correlations} are anti-correlations, with one-sided $p$-values of $\approx 3 \times 10^{-4}$. The first anti-correlation is surprising as both questions 7 and 8 were designed to test the concept of time dilation, and hence were expected to be positively correlated. However, as we shall see in section \ref{Misconceptions}, question 7 (see Table \ref{questions}) is unusual in being one of the two questions having an anti-correlation with confidence. 

There is no obvious relation between the second anti-correlated pair, questions 23 (causality) and 24 (mass-energy). However, question 24 is unusual in being the only question with a negative normalised gain, as can be seen in Fig.~\ref{scores}. Hence we recommend that question 24 be dropped from the RCI.

\subsubsection{Factor analysis}
\label{Factor analysis}

Factor analysis attempts to model students' answers in terms of a small number $T$ of factors, also called latent traits, with $T < K$, the number of questions. In the ideal RCI case these factors would correspond to the nine concepts in Table \ref{concepts} used to design the questions. Factor analysis reproduces the observed data, as accurately as possible, with a linear model of the form \cite{Harman}:
\begin{equation} \label{factor analysis}
M_{ij} = P_i +  \sum_{k=1}^{T} a_{ik} F_{jk} + u_i Y_{ij} ,
\end{equation}
where $M_{ij}$ is the response of student $j$ to question $i$, introduced following Eq.~(\ref{RT model}). The $P_i$ are the item difficulties for each question. The $a_{ik}$ are called the factor loadings. The last term, $u_i Y_{ij}$, is the residual error unique to each question. The $F_{ij}$ and $Y_{ij}$ are independent, normally distributed, random variables with zero mean and unit variance. They represent the underlying larger population from which the data was sampled. Averaging over this population one finds that the factor loadings determine the correlations between questions. Determining these is the primary objective of factor analysis.

The applicability of factor analysis to small sample sizes is controversial. A commonly stated criterion is that meaningful factor analysis requires ten times as many responses as questions \cite{Adams, Scott}. According to this criterion, factor analysis of our data set would not be valid, as we have less than three times as many responses $N$ as questions $K$. 

However Monte Carlo studies have identified more complex criteria that may justify factor analysis of smaller data sets \cite{MacCallum, Mundfrom, deWinter}. Sample sizes as small as ours, $N=63$, may be acceptable if the following three things are all sufficiently high: the number of questions, the ratio of the number of questions to the number of factors \cite{deWinter}, and the factor communalities \cite{MacCallum}. Communalities measure how much of a variable's variance is due to the factor loadings, with a sufficiently high communality in this context being $> 0.6$. The average communality for our post-test questions is $0.74$ \cite{SPSS}. A caveat is that these studies considered continuous data, not binary data like ours. Nevertheless, these studies suggest that under certain conditions a factor analysis of our data may be meaningful, despite the small sample size.

\begin{figure}[tbp]
\includegraphics[width=8cm]{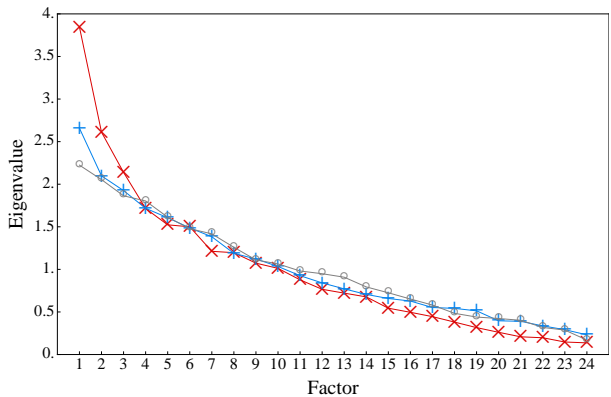}
\caption{(colour online) Scree plots of the eigenvalues resulting from factor analyses of the post-test (red ${\times}$), pre-test (blue +), and random (gray ${\circ}$) data versus the corresponding factor number. The post-test shows four significant eigenvalues, the pre-test two, and the random data none.}
\label{scree plot}
\end{figure}

Figure \ref{scree plot} shows scree plots of the eigenvalues of the question pair correlation matrices. Factor analysis folk lore says that the number of significant factors is the number of eigenvalues on the initial steep slope before the transition to a constant smaller slope. From Fig.~\ref{scree plot} this is four for the post-test data, two for the pre-test data and none for the random data. As mentioned, such low numbers of factors are necessary for the self consistency of our factor analysis \cite{Mundfrom, deWinter}. The random data was generated by a Monte Carlo sampling of all individual question answers with equal probability. It was included as a consistency check that should show no significant factors.

The first four factors for the post-test data have pairs of dominant factor loadings corresponding to the conceptually related pairs in Tables \ref{post-test correlations} and {RT correlations}. In addition, the third factor is dominated by factor loadings for questions 19 and 20 concerning the first postulate concept. The consistency of the factor analysis results with those reported in the previous sections supports its validity. 

\section{Results} 
\label{Results}

The previous section focused on statistical methods and their application to establishing the consistency and reliability of the RCI. In this section the focus is on the implications of the RCI results for special relativity education. We first consider some of the misconceptions revealed by the RCI and then show that the RCI is gender biased.

\subsection{Misconceptions}
\label{Misconceptions}

The RCI confidence scale was briefly described in sections \ref{The Development Process} and \ref{The Students}. From the pre-test to the post-test the average of the correlation between the score and confidence for each question increased from $\langle r_i \rangle = 0.11$ to $\langle r_i \rangle = 0.19$. Most individual questions in the post-test had a positive correlation between confidence and score which indicates some mastery of the relevant concepts. However, two questions had negative correlations: question 7 ($r_7 = -0.3$) and question 23 ($r_{23} = -0.2$), significantly different from zero with $p \lesssim 0.05$.  These negative correlations suggest gaps in students' post-instruction mastery.

Question 7 is given in Table \ref{questions}. It had nearly equal numbers of correct and incorrect answers: item difficulty $P_7 =0.54$. Of those students who rated their confidence as either certain or confident, nearly equal numbers answered correctly and incorrectly. This indicates a misconception about time dilation, which is not captured by the other time dilation questions 5, 6, and 8 that have positive correlations between confidence and score of $r=0.2, 0.25, 0.4$, respectively. One difference between these questions is that the latter are phrased in terms of observations, whereas question 7 is about an experience: travelling across the galaxy. It may be that students are displaying the misconception that while time dilation applies to observations of things, it does not apply to the things themselves.

The other negatively correlated question, 23, is also given in Table \ref{questions}. Most of those who answered it correctly rated their confidence as either guessing or unconfident. Those students may be answering from memorised material, without a firm conceptual  understanding.

Questions 5 and 6 of the RCI, shown in Table \ref{questions}, are a pair testing understanding of time dilation. They ask about the same situation from two different inertial reference frames, with each observer measuring the other's clock to run slow. Their pre-test item difficulties were $P_{5,\textrm{pre}} = 0.63$ and $P_{6,\textrm{pre}} = 0.34$, the difference being significant at the $p=0.05$ level. Furthermore their answers were anti-correlated, $r_{56,\textrm{pre}}= -0.25$, significant at the $p=0.02$ level. 

Correct relativistic thinking would recognise the symmetry between the two reference frames and hence lead to correlation between the answers. However, the anti-correlation suggests an asymmetry misconception in which A measuring B's clock to run slow implies B measuring A's clock to run fast. This is related to absolute motion misconceptions regarding Galilean relativity reported by Panse \textit{et al.} \cite{Panse}. The following student comment from a Real Time Relativity \cite{RTR} lab session on time dilation is an example of both the absolute rest frame and asymmetry misconceptions:

\textit{``The clocks are stationary, and I'm moving  ... so my clock is running slow, which is why the clocks are running fast compared to mine ...''}

As Tables \ref{post-test correlations} and \ref{RT correlations} show, the post-test questions 5 and 6 were the most highly correlated of all pairs, with $r_{56,\textrm{post}}= 0.56$, significant at the $p \le 5\times10^{-5} $ level. This indicates that relativistic thinking has been achieved after instruction, and the asymmetry misconception reduced. The post-test item difficulties were $P_{5,\textrm{post}} = 0.83$ and $P_{6,\textrm{post}} = 0.78$, with corresponding normalised gains of $g_{5} = 0.54$ and $g_{6} = 0.67$.

Evidence from class assessment items indicated that the asymmetry misconception also occurred for length contraction. However, the RCI has no symmetrical pair of length contraction questions to test this. Hence we recommend that a symmetrical partner question be added to the existing RCI length contraction question 13.

\subsection{Gender Differences}
\label{Gender Differences}

In the Physics 2 class we found statistically significant gender differences in the RCI results. The pre-test was taken by 19 females and 51 males, the post-test by 18 females and 45 males. Of those who took both tests 15 were female and 38 were male. As shown in Table \ref{Gender}, males scored higher than females in: the pre-test, post-test, normalised gain, and in confidence. All these differences are significant at the $p \le 0.05$ level according to the Kolmogorov-Smirnov test.

By contrast, the gender groups were statistically identical for assessable homework and for the mid-term exam relativity question. There was also no difference in prior achievement as measured by the ATAR score (discussed in section \ref{The Students}). 

There were only four individual questions for which the gender difference was statistically significant ($p < 0.05$): questions 1 and 2 concerning inertial frames, question 9 concerning velocity addition, and question 17 concerning length contraction. In each of these cases the difference in item difficulty between males and females was $\ge 0.27$. For more than half the questions the magnitude of this difference was $\le 0.1$.

Similar results have been reported for the Force Concept Inventory (FCI) \cite{McCullough, Dietz, Docktor, Coletta} and Brief Electricity and Magnetism Assessment (BEMA) \cite{Kost-Smith}. There is a report of the FCI gender gap being eliminated by high levels of interactive engagement \cite{Lorenzo}, although this has not been found in other studies \cite{Kost}. Other inventories have also been found to have gender differences \cite{Beichner 94, Adams CLASS}. 

Although some authors have claimed that multiple-choice tests are inherently gender biased, the largest studies have found no such effect \cite{HEA, Cole}. 

\begin{table}[t]
\caption{\label{Gender} RCI statistics by gender for the Physics 2 class. $\langle P \rangle$ is the mean item difficulty, $\langle g \rangle$ is the mean normalised gain, $\langle c \rangle$ is the mean confidence,  $\langle x_{\textrm{exam}} \rangle$ is the mean exam score (fraction of possible score) for the students who did the post-test, and $\langle x_{\textrm{hw}} \rangle$ is the mean homework score (fraction of possible score). The ATAR is the university admission score discussed in section \ref{The Students}. $p$-values are the probability that the female and male data were sampled from the same population, so that the observed difference is due to chance.}
\begin{ruledtabular}
\begin{tabular}{cccc}
Statistic &  Females & Males & $p$-value \\ \hline
$\langle P_{\textrm{pre}} \rangle$ & 0.50 & 0.58 & 0.02 \\ 
$\langle P_{\textrm{post}} \rangle$ & 0.63 & 0.72 & 0.003 \\
$\langle g \rangle$ & 0.23 & 0.38 & 0.05 \\
$\langle c_{\textrm{pre}} \rangle$ & 0.41 & 0.53 & 0.02 \\ 
$\langle c_{\textrm{post}} \rangle$ & 0.64 & 0.70 & 0.04 \\ 
$\langle x_{\textrm{exam}} \rangle$ & 0.66 & 0.67 & 0.95 \\
$\langle x_{\textrm{hw}} \rangle$ & 0.75 & 0.75 & 1 \\
$\langle$ATAR$\rangle$ & 94.2 & 93.5 & 0.96 \\
\end{tabular}
\end{ruledtabular}
\end{table}

\section{Conclusions}
\label{Conclusions}

Classical test theory suggests that the RCI may be too easy and, perhaps consequently, insufficiently discriminating. However, we do not recommend revisions, other than those suggested below, until data from a wider range of students has been analysed.

In section \ref{Item response theory} we concluded that question 24, concerning the concept of mass-energy equivalence, should be dropped from the RCI. It has zero discrimination, and is the only question having a negative normalised gain between the pre and post-tests. It was also found to have a strong negative correlation with an apparently unrelated question. If dropped,  the concept of mass-energy equivalence would not be tested by the RCI.

In section \ref{Misconceptions} we concluded that a frame symmetrical pair of length contraction questions is desirable, mirroring the symmetrical pair of time dilation questions. Hence we recommend that a partner question be added to the existing RCI length contraction question 13. However, any such question would require validation along the lines described in sections \ref{The Development Process} and \ref{Data Analysis Methods}.

The evidence presented in section \ref{Gender Differences} suggests that the RCI is gender biased. Previous work has shown similar biases in the Force Concept Inventory and in other concept inventories. Concept inventories are useful because they can help evaluate innovation and hence improve teaching. However if their evaluations are biased with respect to certain student groups there is a risk that improved learning for some comes at the expense of the learning of others. It is a task for future physics education research to investigate and understand this interesting and important problem.

\begin{acknowledgments}
The authors would like to thank the academics and students who helped develop the RCI. We would especially like to acknowledge A.~Wilson for her help with the student interviews, and P.~Francis for his advice on statistical analysis.
\end{acknowledgments}

\begin{widetext}
\pagebreak
\appendix
\section{Supplemental Appendix: The Relativity Concept Inventory}

\maketitle

This is the version of the RCI that was used in the post-test. 
\\
\\

\textbf{Instructions:} 
\begin{itemize}
\item \emph{Some of the questions are multiple choice, with an additional
confidence scale similar to the example below. For each of these questions,
circle the answer that you agree most with, and mark on the scale
how confident you are in your choice. }
\end{itemize}
\begin{eqnarray*}
 & \text{Rate how confident you are in your answer:}\\
 & \bigcirc\cdots\cdots\cdots\bigcirc\cdots\cdots\cdots\bigcirc\cdots\cdots\cdots\bigcirc\cdots\cdots\cdots\bigcirc\\
 & \text{guessing\quad\quad unconfident\quad\ \ neutral\quad\quad confident\quad\quad\ certain\quad}
\end{eqnarray*}

\begin{itemize}
\item \emph{Some of the questions are in the form of statements with which
you may agree or disagree. Circle the response that most closely corresponds
to your position on the question.}
\item \emph{In all of the following questions, the symbol $c$ represents
the speed of light in a vacuum, $3\times10^{8}\text{ m/s}$.}
\item \emph{Answer all of the questions to the best of your knowledge.}\\
\emph{\pagebreak{}}
\end{itemize}
In the following two questions, Alice is standing in a train moving
at velocity $v$ from \textbf{left to right} relative to Bob, who
is standing on a platform. As Alice passes Bob, she drops a bowling
ball out of the train's window:

\begin{center}
\includegraphics[scale=0.3]{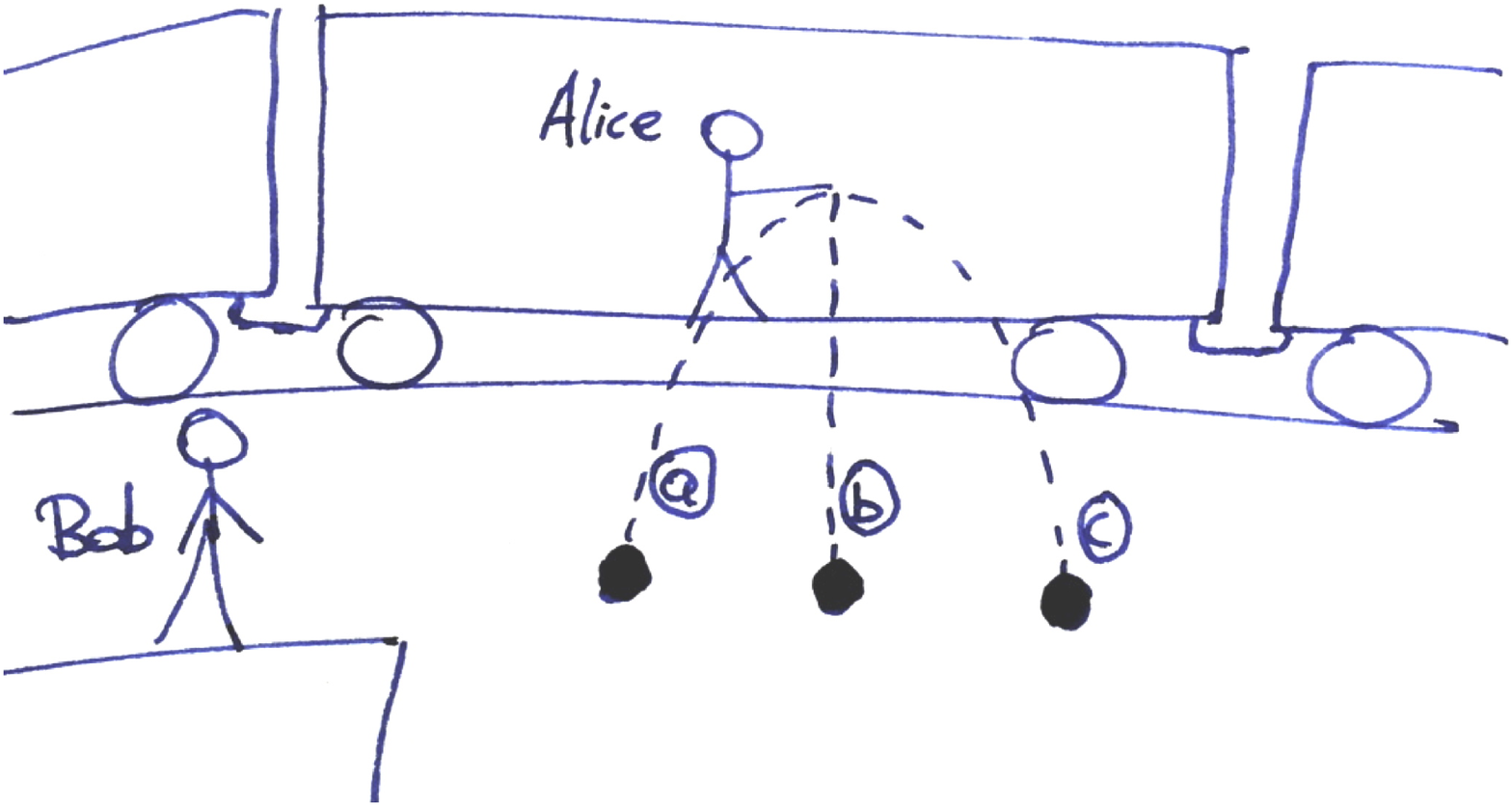}
\par\end{center}
\begin{enumerate}
\item Ignoring air resistance, which path of the ball would Bob observe,
standing on the platform?

\begin{enumerate}
\item Path (a)
\item Path (b)
\item Path (c)
\end{enumerate}
\end{enumerate}
\begin{eqnarray*}
 & \text{Rate how confident you are in your answer:}\\
 & \bigcirc\cdots\cdots\cdots\bigcirc\cdots\cdots\cdots\bigcirc\cdots\cdots\cdots\bigcirc\cdots\cdots\cdots\bigcirc\\
 & \text{guessing\quad\quad unconfident\quad\ \ neutral\quad\quad confident\quad\quad\ certain\quad}
\end{eqnarray*}

\vspace{0.5 cm}

\begin{enumerate}
\item [2.]Ignoring air resistance, which path of the ball would Alice observe,
standing in the train?

\begin{enumerate}
\item Path (a)
\item Path (b)
\item Path (c)
\end{enumerate}
\end{enumerate}
\begin{eqnarray*}
 & \text{Rate how confident you are in your answer:}\\
 & \bigcirc\cdots\cdots\cdots\bigcirc\cdots\cdots\cdots\bigcirc\cdots\cdots\cdots\bigcirc\cdots\cdots\cdots\bigcirc\\
 & \text{guessing\quad\quad unconfident\quad\ \ neutral\quad\quad confident\quad\quad\ certain\quad}
\end{eqnarray*}

\vspace{0.5 cm}

\begin{enumerate}
\item [3.]True or false: ``In principle, it is possible for an observer
following a pulse of light at a constant high speed to observe the
light to be almost stationary.''

\begin{enumerate}
\item True
\item False
\end{enumerate}
\end{enumerate}
\begin{eqnarray*}
 & \text{Rate how confident you are in your answer:}\\
 & \bigcirc\cdots\cdots\cdots\bigcirc\cdots\cdots\cdots\bigcirc\cdots\cdots\cdots\bigcirc\cdots\cdots\cdots\bigcirc\\
 & \text{guessing\quad\quad unconfident\quad\ \ neutral\quad\quad confident\quad\quad\ certain\quad}
\end{eqnarray*}

\vspace{0.5 cm}

\begin{enumerate}
\item [4.]Consider a spaceship travelling from Earth towards a distant
star at a constant high velocity $v$ relative to Earth. The spaceship
sends a light pulse back to Earth. On Earth, the speed of this pulse
is measured to be:

\begin{enumerate}
\item $c$
\item $c+v$
\item $c-v$
\end{enumerate}
\end{enumerate}
\begin{eqnarray*}
 & \text{Rate how confident you are in your answer:}\\
 & \bigcirc\cdots\cdots\cdots\bigcirc\cdots\cdots\cdots\bigcirc\cdots\cdots\cdots\bigcirc\cdots\cdots\cdots\bigcirc\\
 & \text{guessing\quad\quad unconfident\quad\ \ neutral\quad\quad confident\quad\quad\ certain\quad}
\end{eqnarray*}

\vspace{0.5 cm}

In the following two questions, Abbey is in a spaceship moving at
high speed relative to Brendan, who is standing on an asteroid (a
very small piece of rock floating in space). She flies past him so
that at $t=0$, she is momentarily adjacent to Brendan.
\begin{enumerate}

\vspace{0.5 cm}

\item [5.]At the instant that Abbey's ship passes Brendan, she sends two
light pulses to him from her ship. If the light pulses are emitted
a nanosecond ($10^{-9}$ seconds) apart according to Abbey's clock,
what will be the time interval between the pulses according to Brendan?

\begin{enumerate}
\item Greater than one nanosecond
\item Equal to one nanosecond
\item Less than one nanosecond
\end{enumerate}
\end{enumerate}
\begin{eqnarray*}
 & \text{Rate how confident you are in your answer:}\\
 & \bigcirc\cdots\cdots\cdots\bigcirc\cdots\cdots\cdots\bigcirc\cdots\cdots\cdots\bigcirc\cdots\cdots\cdots\bigcirc\\
 & \text{guessing\quad\quad unconfident\quad\ \ neutral\quad\quad confident\quad\quad\ certain\quad}
\end{eqnarray*}

\vspace{0.5 cm}

\begin{enumerate}
\item [6.]Also while Abbey's ship passes Brendan, Brendan sends two light
pulses to Abbey. If Brendan sends the light pulses a nanosecond ($10^{-9}$
seconds) apart according to his clock, what will be the time interval
between the pulses according to Abbey?

\begin{enumerate}
\item Greater than one nanosecond
\item Equal to one nanosecond
\item Less than one nanosecond
\end{enumerate}
\end{enumerate}
\begin{eqnarray*}
 & \text{Rate how confident you are in your answer:}\\
 & \bigcirc\cdots\cdots\cdots\bigcirc\cdots\cdots\cdots\bigcirc\cdots\cdots\cdots\bigcirc\cdots\cdots\cdots\bigcirc\\
 & \text{guessing\quad\quad unconfident\quad\ \ neutral\quad\quad confident\quad\quad\ certain\quad}
\end{eqnarray*}

\vspace{0.5 cm}

\begin{enumerate}
\item [7.]It is known that our galaxy is of the order of $100,000$ light-years
in diameter. True or false: ``Travelling at a constant speed that
is less than, but close to, the speed of light, in principle it is
possible for a person to cross the galaxy within their lifetime.''

\begin{enumerate}
\item True
\item False
\end{enumerate}
\end{enumerate}
\begin{eqnarray*}
 & \text{Rate how confident you are in your answer:}\\
 & \bigcirc\cdots\cdots\cdots\bigcirc\cdots\cdots\cdots\bigcirc\cdots\cdots\cdots\bigcirc\cdots\cdots\cdots\bigcirc\\
 & \text{guessing\quad\quad unconfident\quad\ \ neutral\quad\quad confident\quad\quad\ certain\quad}
\end{eqnarray*}

\pagebreak{}

\begin{enumerate}
\item [8.]The Olympic Games is a two-week long sports competition. An interested
alien astronomer watches the Olympics from a distant planet moving
at high speed relative to Earth. \emph{If the alien were to compensate
for the time the light from Earth takes to reach them}, they would
measure the length of the Olympics to be:

\begin{enumerate}
\item Greater than two weeks
\item Equal to two weeks
\item Less than two weeks
\end{enumerate}
\end{enumerate}
\begin{eqnarray*}
 & \text{Rate how confident you are in your answer:}\\
 & \bigcirc\cdots\cdots\cdots\bigcirc\cdots\cdots\cdots\bigcirc\cdots\cdots\cdots\bigcirc\cdots\cdots\cdots\bigcirc\\
 & \text{guessing\quad\quad unconfident\quad\ \ neutral\quad\quad confident\quad\quad\ certain\quad}
\end{eqnarray*}

\vspace{0.5 cm}

In the following two questions, the scenario is as follows: Alex and
his friend Bianca decide to set off on separate voyages in identical
spaceships. They each speed away from Earth in opposite directions
- Alex at $v=0.75c$ to the left, and Bianca at $v=0.75c$ to the
right, relative to an observer on Earth. 
\begin{enumerate}
\item [9.]If Alex measures the rate at which his distance to Bianca is
increasing, he will obtain a value that is:

\begin{enumerate}
\item Equal to $1.5c$
\item Greater than $c$ but less than $1.5c$
\item Equal to $c$
\item Greater than $0.75c$ but less than $c$
\item Equal to $0.75c$
\end{enumerate}
\end{enumerate}
\begin{eqnarray*}
 & \text{Rate how confident you are in your answer:}\\
 & \bigcirc\cdots\cdots\cdots\bigcirc\cdots\cdots\cdots\bigcirc\cdots\cdots\cdots\bigcirc\cdots\cdots\cdots\bigcirc\\
 & \text{guessing\quad\quad unconfident\quad\ \ neutral\quad\quad confident\quad\quad\ certain\quad}
\end{eqnarray*}

\vspace{0.5 cm}

\begin{enumerate}
\item [10.]If Cameron, an observer on Earth, measures the rate at which
the distance between Alex and Bianca is increasing, he will obtain
a value that is:

\begin{enumerate}
\item Equal to $1.5c$
\item Greater than $c$ but less than $1.5c$
\item Equal to $c$
\item Greater than $0.75c$ but less than $c$
\item Equal to $0.75c$
\end{enumerate}
\end{enumerate}
\begin{eqnarray*}
 & \text{Rate how confident you are in your answer:}\\
 & \bigcirc\cdots\cdots\cdots\bigcirc\cdots\cdots\cdots\bigcirc\cdots\cdots\cdots\bigcirc\cdots\cdots\cdots\bigcirc\\
 & \text{guessing\quad\quad unconfident\quad\ \ neutral\quad\quad confident\quad\quad\ certain\quad}
\end{eqnarray*}

\pagebreak{}

In the following four questions, Amanda is standing on a train travelling
at high speed past Bryan, who is standing on a platform. As she passes
Bryan, she drops two bowling balls out of the window at the same time
(Amanda's time), and from an arm's span apart. 

\begin{center}
\includegraphics[scale=0.4]{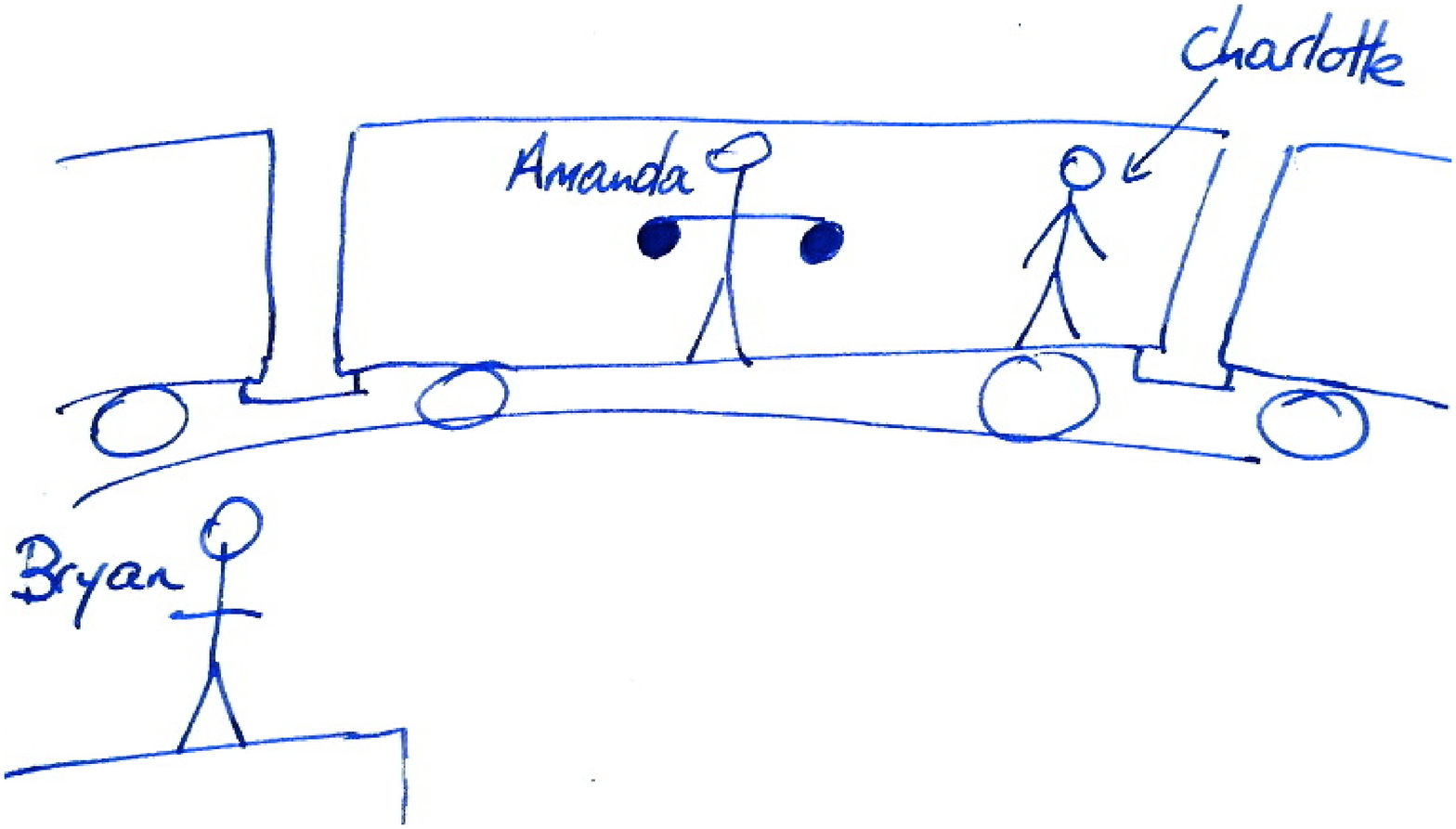}
\par\end{center}
\begin{enumerate}
\item [11.]Bryan stands on the platform and watches the balls fall to the
ground. \emph{If he compensates for the time that the light from the
impacts takes to reach him}, in what order does Bryan measure the
balls hitting the ground?

\begin{enumerate}
\item At the same time
\item One ball before the other
\end{enumerate}
\end{enumerate}
\begin{eqnarray*}
 & \text{Rate how confident you are in your answer:}\\
 & \bigcirc\cdots\cdots\cdots\bigcirc\cdots\cdots\cdots\bigcirc\cdots\cdots\cdots\bigcirc\cdots\cdots\cdots\bigcirc\\
 & \text{guessing\quad\quad unconfident\quad\ \ neutral\quad\quad confident\quad\quad\ certain\quad}
\end{eqnarray*}

\vspace{0.5 cm}

\begin{enumerate}
\item [12.]Charlotte is another passenger on the train with Amanda. \emph{If
she compensates for the time that the light from the impacts takes
to reach her}, in what order does Charlotte measure the balls hitting
the ground?

\begin{enumerate}
\item At the same time
\item One ball before the other
\end{enumerate}
\end{enumerate}
\begin{eqnarray*}
 & \text{Rate how confident you are in your answer:}\\
 & \bigcirc\cdots\cdots\cdots\bigcirc\cdots\cdots\cdots\bigcirc\cdots\cdots\cdots\bigcirc\cdots\cdots\cdots\bigcirc\\
 & \text{guessing\quad\quad unconfident\quad\ \ neutral\quad\quad confident\quad\quad\ certain\quad}
\end{eqnarray*}

\vspace{0.5 cm}

\begin{enumerate}
\item [13.]Amanda has an arm span of $D$ meters at rest. If Bryan performs
a measurement of Amanda's arm span as she passes him, he will obtain
a value:

\begin{enumerate}
\item Greater than $D$
\item Equal to $D$
\item Less than $D$
\end{enumerate}
\end{enumerate}
\begin{eqnarray*}
 & \text{Rate how confident you are in your answer:}\\
 & \bigcirc\cdots\cdots\cdots\bigcirc\cdots\cdots\cdots\bigcirc\cdots\cdots\cdots\bigcirc\cdots\cdots\cdots\bigcirc\\
 & \text{guessing\quad\quad unconfident\quad\ \ neutral\quad\quad confident\quad\quad\ certain\quad}
\end{eqnarray*}

\begin{enumerate}
\item [14.]Amanda also has a height of $H$ meters at rest. If Bryan performs
a measurement of Amanda's height as she passes him, he will obtain
a value:

\begin{enumerate}
\item Greater than $H$
\item Equal to $H$
\item Less than $H$
\end{enumerate}
\end{enumerate}
\begin{eqnarray*}
 & \text{Rate how confident you are in your answer:}\\
 & \bigcirc\cdots\cdots\cdots\bigcirc\cdots\cdots\cdots\bigcirc\cdots\cdots\cdots\bigcirc\cdots\cdots\cdots\bigcirc\\
 & \text{guessing\quad\quad unconfident\quad\ \ neutral\quad\quad confident\quad\quad\ certain\quad}
\end{eqnarray*}

\vspace{0.5 cm}

\begin{enumerate}
\item [15.]Two separate light bulbs emit flashes of light, distant from
an observer. This observer receives the light from both flashes at
the same time. From this alone it is possible to conclude that:

\begin{enumerate}
\item The flashes occurred at the same time for all observers
\item The flashes occurred at the same time for the observer at that location
\item The flashes occurred at the same time if the observer is not moving
relative to the light bulbs
\item It is not possible to make any of the above conclusions
\end{enumerate}
\end{enumerate}
\begin{eqnarray*}
 & \text{Rate how confident you are in your answer:}\\
 & \bigcirc\cdots\cdots\cdots\bigcirc\cdots\cdots\cdots\bigcirc\cdots\cdots\cdots\bigcirc\cdots\cdots\cdots\bigcirc\\
 & \text{guessing\quad\quad unconfident\quad\ \ neutral\quad\quad confident\quad\quad\ certain\quad}
\end{eqnarray*}

\vspace{0.5 cm}

\begin{enumerate}
\item [16.]In the following thought experiment, you are in a high speed
train travelling along a railway. True or false: ``If you measure
the dimensions of the train compartment, you will obtain different
values than if the train were at rest.''

\begin{enumerate}
\item True
\item False
\end{enumerate}
\end{enumerate}
\begin{eqnarray*}
 & \text{Rate how confident you are in your answer:}\\
 & \bigcirc\cdots\cdots\cdots\bigcirc\cdots\cdots\cdots\bigcirc\cdots\cdots\cdots\bigcirc\cdots\cdots\cdots\bigcirc\\
 & \text{guessing\quad\quad unconfident\quad\ \ neutral\quad\quad confident\quad\quad\ certain\quad}
\end{eqnarray*}

\vspace{0.5 cm}

\begin{enumerate}
\item [17.]Consider a futuristic space station that specialises in constructing
fast spaceships. Once the ships are built, they leave the station
at high speed for testing. As they leave the station at speed, a serial
number is stamped instantaneously on the side of the ship by a machine
on the station. This serial number has length $D$ as measured by
a builder on the space station. After the ship has finished its test
run, it returns to the station and is parked in the garage. What is
the length of the serial number now, as measured by the builder on
the space station?

\begin{enumerate}
\item Greater than $D$
\item Equal to $D$
\item Less than $D$
\end{enumerate}
\end{enumerate}
\begin{eqnarray*}
 & \text{Rate how confident you are in your answer:}\\
 & \bigcirc\cdots\cdots\cdots\bigcirc\cdots\cdots\cdots\bigcirc\cdots\cdots\cdots\bigcirc\cdots\cdots\cdots\bigcirc\\
 & \text{guessing\quad\quad unconfident\quad\ \ neutral\quad\quad confident\quad\quad\ certain\quad}
\end{eqnarray*}

\begin{enumerate}
\item [18.]Adam is in a spaceship moving at $v=0.99c$ relative to our
galaxy. Adam wants to measure the mass of his ship by observing how
resistant the ship is to acceleration. If Adam exerts a force on the
ship (by turning on a rocket engine, for example) and measures (with
an accelerometer inside the ship) the acceleration that results, he
will obtain a value that is:

\begin{enumerate}
\item Greater than what he would measure if his ship were at rest relative
to the galaxy.
\item Equal to what he would measure if his ship were at rest relative to
the galaxy.
\item Less than what he would measure if his ship were at rest relative
to the galaxy.
\end{enumerate}
\end{enumerate}
\begin{eqnarray*}
 & \text{Rate how confident you are in your answer:}\\
 & \bigcirc\cdots\cdots\cdots\bigcirc\cdots\cdots\cdots\bigcirc\cdots\cdots\cdots\bigcirc\cdots\cdots\cdots\bigcirc\\
 & \text{guessing\quad\quad unconfident\quad\ \ neutral\quad\quad confident\quad\quad\ certain\quad}
\end{eqnarray*}

\vspace{0.5 cm}

\begin{enumerate}
\item [19.]In the following thought experiment, you are in a high speed
train travelling along a railway. True or false: ``If you measure
the rate at which your watch is ticking, you will obtain a different
value than if the train were at rest.''

\begin{enumerate}
\item True
\item False
\end{enumerate}
\end{enumerate}
\begin{eqnarray*}
 & \text{Rate how confident you are in your answer:}\\
 & \bigcirc\cdots\cdots\cdots\bigcirc\cdots\cdots\cdots\bigcirc\cdots\cdots\cdots\bigcirc\cdots\cdots\cdots\bigcirc\\
 & \text{guessing\quad\quad unconfident\quad\ \ neutral\quad\quad confident\quad\quad\ certain\quad}
\end{eqnarray*}

\vspace{0.5 cm}

\begin{enumerate}
\item [20.]You are in a well equipped physics lab without windows or ways
of interacting with the outside world. It is known that the lab is
in uniform motion. How do you determine the velocity of the lab?

\begin{enumerate}
\item You throw a ball across the lab and measure its change in velocity
\item You shine a laser beam across the lab and measure its change in velocity
\item Either (a) or (b)
\item It is not possible to determine the lab's velocity by experiment
\end{enumerate}
\end{enumerate}
\begin{eqnarray*}
 & \text{Rate how confident you are in your answer:}\\
 & \bigcirc\cdots\cdots\cdots\bigcirc\cdots\cdots\cdots\bigcirc\cdots\cdots\cdots\bigcirc\cdots\cdots\cdots\bigcirc\\
 & \text{guessing\quad\quad unconfident\quad\ \ neutral\quad\quad confident\quad\quad\ certain\quad}
\end{eqnarray*}

\pagebreak{}

\begin{enumerate}
\item [21.]You observe a set of distant, spatially separated clocks that
are synchronised in their rest frame. You are at rest relative to
the clocks, and you observe (through a telescope) that the times read
on the clocks are different. This is due to:

\begin{enumerate}
\item Time dilation
\item Length contraction
\item Relativity of simultaneity
\item None of the above
\end{enumerate}
\end{enumerate}
\begin{eqnarray*}
 & \text{Rate how confident you are in your answer:}\\
 & \bigcirc\cdots\cdots\cdots\bigcirc\cdots\cdots\cdots\bigcirc\cdots\cdots\cdots\bigcirc\cdots\cdots\cdots\bigcirc\\
 & \text{guessing\quad\quad unconfident\quad\ \ neutral\quad\quad confident\quad\quad\ certain\quad}
\end{eqnarray*}

\vspace{0.5 cm}

\begin{enumerate}
\item [22.]If two events are separated in such a way that an observer can
be present at both events, which relationship(s) between the two events
are the same for all observers? 

\begin{enumerate}
\item The time between the two events
\item The distance between the two events
\item The order in which the events occur
\item None of these relationships are the same for all observers
\end{enumerate}
\end{enumerate}
\begin{eqnarray*}
 & \text{Rate how confident you are in your answer:}\\
 & \bigcirc\cdots\cdots\cdots\bigcirc\cdots\cdots\cdots\bigcirc\cdots\cdots\cdots\bigcirc\cdots\cdots\cdots\bigcirc\\
 & \text{guessing\quad\quad unconfident\quad\ \ neutral\quad\quad confident\quad\quad\ certain\quad}
\end{eqnarray*}

\vspace{0.5 cm}

\begin{enumerate}
\item [23.]If two events are separated in such a way that \textbf{no} observer
can be present at both events, which relationship(s) between the two
events are the same for all observers?

\begin{enumerate}
\item The time between the two events
\item The distance between the two events
\item The order in which the events occur
\item None of these relationships are the same for all observers
\end{enumerate}
\end{enumerate}
\begin{eqnarray*}
 & \text{Rate how confident you are in your answer:}\\
 & \bigcirc\cdots\cdots\cdots\bigcirc\cdots\cdots\cdots\bigcirc\cdots\cdots\cdots\bigcirc\cdots\cdots\cdots\bigcirc\\
 & \text{guessing\quad\quad unconfident\quad\ \ neutral\quad\quad confident\quad\quad\ certain\quad}
\end{eqnarray*}

\vspace{0.5 cm}

\begin{enumerate}
\item [24.]Consider a closed box, containing an equal amount of matter
and antimatter. The total mass of this box and its contents is initially
$M$. The matter and antimatter are then allowed to annihilate inside
the box, turning into photons in the process. What is the total mass
of the box and its contents \emph{after} the annihilation?

\begin{enumerate}
\item Greater than $M$
\item Equal to $M$
\item Less than $M$
\end{enumerate}
\end{enumerate}
\begin{eqnarray*}
 & \text{Rate how confident you are in your answer:}\\
 & \bigcirc\cdots\cdots\cdots\bigcirc\cdots\cdots\cdots\bigcirc\cdots\cdots\cdots\bigcirc\cdots\cdots\cdots\bigcirc\\
 & \text{guessing\quad\quad unconfident\quad\ \ neutral\quad\quad confident\quad\quad\ certain\quad}
\end{eqnarray*}

\begin{center}
\rule[0.5ex]{0.25\columnwidth}{1pt}
\par\end{center}

\end{widetext}


\begin{thebibliography}{99}

\bibitem{Inventory list} Assessment Instrument Information Page, NC State Physics Education R and D Group, http://www.ncsu.edu/per/TestInfo.html, accessed February 26, 2013.

\bibitem{Wegener} M. Wegener, T. McIntyre, D. McGrath, C. Savage, M. Williamson, Developing a Virtual Physics World, Australasian Journal of Educational Technology, \textbf{28} (Special issue, 3), 504 (2012).

\bibitem{McGrath} D. McGrath, M. Wegener, T. McIntyre, C. Savage, M. Williamson, Student experiences of virtual reality - a case study in learning special relativity, Am. J. Phys. \textbf{78}, 862 (2010).

\bibitem{Savage} C. M. Savage, A. Searle, L. McCalman, Real Time Relativity: exploratory learning of special relativity,  Am. J. Phys. \textbf{75}, 791 (2007).

\bibitem{Hewson 1982} Peter W. Hewson, A case study of conceptual change in special relativity: the influence of prior knowledge in learning, Eur. J. Sci. Educ. \textbf{4}, 61 (1982).

\bibitem{Posner 1982} George J. Posner, Kenneth A. Strike, Peter W. Hewson, and William A. Gertzog, Accommodation of a scientific conception: Toward a theory of conceptual change, Sci. Educ. \textbf{2}, 211 (1982).

\bibitem{Villani 1987} A. Villani, and J. L. A. Pacca, Students' spontaneous ideas about the speed of light, Int. J. Sci. Educ. \textbf{9}, 55 (1987).

\bibitem{Mermin 1994} N. D. Mermin, Lapses in relativistic pedagogy, Am. J. Phys. {\bf 62}, 11 (1994).

\bibitem{Scherr 2001} R. E. Scherr, P. S. Shaffer, and S. Vokos, Student understanding of time in special relativity: simultaneity and reference frames, Phys. Educ. Res., Am. J. Phys. Suppl.  {\bf 69}, S24 (2001).

\bibitem{Scherr 2002} R. E. Scherr, P. S. Shaffer, and S. Vokos, The challenge of changing deeply held student beliefs about the relativity of simultaneity, Am. J. Phys. {\bf 70}, 1238 (2002).

\bibitem{Belloni 2004} M. Belloni, W. Christian, and M. H. Darcy, Teaching special relativity using physlets, Phys. Teach. {\bf 42}, 8 (2004).

\bibitem{Scherr 2007} R. E. Scherr, Modeling student thinking: An example from special relativity, Am. J. Phys. {\bf 75}, 272 (2007).

\bibitem{Chasteen} Stephanie V. Chasteen, Rachel E. Pepper, Marcos D. Caballero, Steven J. Pollock, and Katherine K. Perkins, Colorado Upper-Division Electrostatics diagnostic: A conceptual assessment for the junior level, Phys. Rev. ST Phys. Educ. Res. \textbf{8}, 020108 (2012).

\bibitem{Adams} Wendy K. Adams and Carl E. Wieman, Development and Validation of Instruments to Measure Learning of Expert-Like Thinking, Int. J. Sci. Educ. \textbf{33}, 1289 (2011).

\bibitem{Gibson 2008} K. Gibson, Special relativity in the classroom, doctoral thesis, Arizona State University (2008).

\bibitem{Texts} Edwin F. Taylor and John Archibald Wheeler, Spacetime physics: introduction to special relativity, W. H. Freeman \& Co., New York, 1982;  Nathaniel David Mermin, It's about time, Princeton University Press, New Jersey, 2005; Sean Carroll, Spacetime and geometry, Addison-Wesley, San Francisco, 2004; Wolfgang Rindler, Relativity, special, general and cosmological, Oxford University Press, 2006.

\bibitem{experts} Expert input was solicited by email from the members of the Australasian Society for General Relativity and Gravitation, and from members of the Education Group of the Australian Institute of Physics. Responses were also obtained through a posting to the Matter and Interactions Yahoo group.

\bibitem{Allen} Kirk Allen, Teri Reed-Rhoads, and Robert Terry. Work in progress: Assessing student confidence of introductory statistics concepts. In: Proceedings - Frontiers in Education - 36th Annual Conference, pgs. 13-14. IEEE (2006).

\bibitem{Allen 2004} Kirk Allen, Andrea Stone, Teri Reed-Rhoads, and Teri J. Murphy, The Statistics Concepts Inventory: Developing a valid and reliable instrument, Proceedings, American Society for Engineering Education Annual Conference \& Exposition (2004).

\bibitem{Allen 2006} Kirk Allen, Teri Reed-Rhoads, and Terry Robert, Work in progress: Assessing student confidence of introductory statistics concepts, Proceedings. Frontiers in Education. 36th Annual Conference (2006).

\bibitem{Sharma} Manjula Devi Sharma and James Bewes. Self-monitoring: Confidence, academic achievement and gender differences in physics. Journal of Learning Design \textbf{4}, issue 3 (2011).

\bibitem{ethics} This project was approved by the ANU Human Research Ethics Committee: human ethics protocol 2012/380.

\bibitem{RTR} Real Time Relativity website, http://realtimerelativity.org, accessed 26 February 2013.

\bibitem{ATAR} The ATAR score ranks graduating secondary school students. The Physics 2 median ATAR score of 95 indicates that the median student scored higher than 95\% of other year 12 students.

\bibitem{Silver} Nate Silver. The Signal and the Noise. The Penguin Press, New Yory, 2012. Chapters 5 and 8.

\bibitem{numerical recipes} William H. Press, Brian P. Flannery, Saul A. Teukolsky, William T. Vetterling. Numerical Recipes: the art of scientific computing. Cambridge University Press, New York, 1st edition, 1986.

\bibitem{Wackerly}  Dennis D. Wackerly, William III Mendenhall, and Richard L. Scheaffer. Mathematical Statistics with Applications. Thomson, Belmont, 7th edition, 2008.

\bibitem{Ding2006} Lin Ding, Ruth Chabay, Bruce Sherwood, and Robert Beichner, Evaluating an electricity and magnetism assessment tool: Brief electricity and magnetism assessment, Phys. Rev. ST Phys. Educ. Res. \textbf{2}, 010105 (2006).

\bibitem{Ding} Lin Ding and Robert Beichner, Approaches to data analysis of multiple-choice questions, Phys. Rev. ST Phys. Educ. Res. \textbf{5}, 020103 (2009).

\bibitem{Hake 98} Richard R. Hake, Interactive-engagement versus traditional methods: A six-thousand-student survey of mechanics test data for introductory physics courses, Am. J. Phys. {\bf 66}, 64 (1998).

\bibitem{Mathematica} We used the MultiNomialDistribution function of Mathematica 8 to generate our random samples.

\bibitem{Francis} This idea was suggested to us by Dr. Paul Francis of ANU.

\bibitem{Planinic} Maja Planinic, Lana Ivanjek, and Ana Susac, Rasch model based analysis of the Force Concept Inventory, Phys. Rev. ST Phys. Educ. Res. \textbf{6}, 010103 (2010).

\bibitem{ministeps} We used Ministep to do the Rasch analysis: http://www.winsteps.com/ministep.htm, accessed 23 February 2013.

\bibitem{Harman} Harry H. Harman. Modern Factor Analysis. University of Chicago Press, Chicago, 3rd edition 1976.

\bibitem{Scott} Terry F. Scott, Daniel Schumayer, and Andrew R. Gray, Exploratory factor analysis of a Force Concept Inventory data set, Phys. Rev. ST Phys. Educ. Res. \textbf{8}, 020105 (2012).

\bibitem{MacCallum} Robert C. MacCallum, Keith F. Widaman, Shaobo Zhang, and Sehee Hong, Sample size in factor analysis, Psychological Methods \textbf{4}, 84 (1999).

\bibitem{Mundfrom} Daniel J. Mundfrom, Dale G. Shaw, and Tian Lu Ke, Minimum Sample Size Recommendations for Conducting Factor Analyses, Int. J. Testing \textbf{5}, 159 (2005).

\bibitem{deWinter} J.C.F. de Winter, D. Dodou, and P. A. Wieringa, Exploratory factor analysis with small sample sizes, Multivariate Behavioral Research \textbf{44}, 147 (2009).

\bibitem{SPSS} We used the statistical analysis package SPSS for our factor analysis. It calculates question communalities.

\bibitem{Panse} Sudhir Panse, Jayashree Ramadas, and Arvind Kumar, Alternative conceptions in Galilean relativity: frames of reference, Int. J. Sci. Educ. \textbf{16}, 63 (1994).

\bibitem{McCullough}  Laura McCullough, Gender, context and physics assessment, J. Int. Women's Studies, \textbf{5}, 20 (2004).

\bibitem{Dietz} R. D. Dietz, R. H. Pearson, M. R. Semak, and C. W. Willis, Gender bias in the Force Concept Inventory? In, AIP Conference Proceedings 1413: Physics Education Research Conference, pgs. 171-174 (2011).

\bibitem{Docktor} Jennifer Docktor and Kenneth Heller, Gender Differences in Both Force Concept Inventory and Introductory Physics Performance. In, AIP Conference Proceedings 1064: Physics Education Research Conference, pgs. 15-18 (2008).

\bibitem{Coletta} Vincent P. Coletta, Jeffery Phillips, and Jeffery J. Steinert, FCI normalized gain, scientific reasoning ability, thinking in physics, and gender effects. In, AIP Conference Proceedings 1413: Physics Education Research Conference, pgs. 23-26 (2011).

\bibitem{Kost-Smith} Lauren E. Kost-Smith, Steven J. Pollock and Noah D. Finkelstein, Gender disparities in second-semester college physics: the incremental effects of a smog of bias, Phys. Rev. ST Phys. Educ. Res. \textbf{6}, 020112 (2010).

\bibitem{Lorenzo} Mercedes Lorenzo, Catherine H. Crouch, and Eric Mazur, Reducing the gender gap in the physics classroom, Am. J. Phys. {\bf 74}, 118 (2006).

\bibitem{Kost} L. E. Kost, S. J. Pollock and N. D. Finkelstein, Characterizing the gender gap in introductory physics, Phys. Rev. ST Phys. Educ. Res. \textbf{5}, 010101 (2009).

\bibitem{Beichner 94} Robert J. Beichner, Testing student interpretation of kinematics graphs, Am. J. Phys. {\bf 62}, 750 (1994).

\bibitem{Adams CLASS} W. K. Adams, K. K. Perkins, N. S. Podolefsky, M. Dubson, N. D. Finkelstein, and C. E. Wieman, New instrument for measuring student beliefs about physics and learning physics: The Colorado Learning Attitudes about Science Survey, Phys. Rev. ST Phys. Educ. Res. \textbf{2}, 010101 (2006).

\bibitem{HEA} J. McKendree and C. Smith, FAQ - I have heard that multiple-choice questions (MCQs) are biased in favour of males. What is the evidence for this?,  http://www.heacademy.ac.uk/resources/detail/subjects/ \newline medev/FAQ-I-have-heard-that-multiple-choice-questions-are-biased-in-favour-of-males, accessed 23 February 2013.

\bibitem{Cole} N. Cole, The ETS Gender Study: How Females and Males Perform in Educational Settings. Educational Testing Service Technical Report (1997), http://www.eric.ed.gov/ERICWebPortal/detail?accno= \newline ED424337, accessed 23 February 2013.

\end{thebibliography}
\end{document}